\documentclass[11pt]{article}

\usepackage[preprint,nonatbib]{nips_gibs}

\usepackage[utf8]{inputenc} % allow utf-8 input
\usepackage[T1]{fontenc}    % use 8-bit T1 fonts
\usepackage{url}            % simple URL typesetting
\usepackage{booktabs}       % professional-quality tables
\usepackage{amsfonts}       % blackboard math symbols
\usepackage{nicefrac}       % compact symbols for 1/2, etc.
\usepackage{microtype}      % microtypography

\usepackage[english]{babel}

\usepackage{amsmath}
\usepackage{amssymb}
\usepackage{amsthm}

\usepackage{graphicx}

\usepackage{enumitem}
\setitemize{noitemsep,topsep=2pt,parsep=2pt,partopsep=2pt}
\setenumerate{noitemsep,topsep=2pt,parsep=2pt,partopsep=2pt}

\usepackage[parfill]{parskip}
\usepackage{tabularx}
\renewcommand{\arraystretch}{1.1}

\usepackage{subcaption}

\newcommand{\DM} {\normalfont{\textsf{DM}}}
\newcommand{\dDM} {\normalfont{\textsf{dDM}}}
\newcommand{\dPDM} {\normalfont{\textsf{dPDM}}}
\newcommand{\Hash}{\mathsf{Hash}}
\newcommand{\AUX}{\mathsf{Aux}}

\newcommand{\fraB}{\ensuremath{\mathcal{B}}}
\newcommand{\fraN}{\ensuremath{\mathcal{N}}}
\newcommand{\fraS}{\ensuremath{\mathcal{S}}}
\newcommand{\fraT}{\ensuremath{\mathcal{T}}}

\newcommand{\Z}{\ensuremath{\mathbb{Z}}}

\theoremstyle{definition}

\begin{document}

\title{WibsonTree: Efficiently Preserving Seller's Privacy in a Decentralized Data Marketplace}

\author{Ariel Futoransky \\
	Disarmista
	\And Carlos Sarraute \\
	Wibson \& Grandata
	\And Ariel Waissbein \\
	Disarmista
	\AND Matias Travizano \\
	Wibson
	\And Daniel Fernandez \\
	Wibson 
}

\maketitle

\begin{abstract}

We present a cryptographic primitive called \emph{WibsonTree} designed to preserve users' privacy by allowing them to demonstrate predicates on their personal attributes, without revealing the values of those attributes. We suppose that there are three types of agents ---buyers, sellers and notaries--- who interact in a decentralized privacy-preserving data marketplace (\dPDM) such as the Wibson marketplace. We introduce the \emph{WibsonTree} protocol as an efficient cryptographic primitive that enables the exchange of private information while preserving the seller's privacy. Using our primitive, a data seller can efficiently prove that he/she belongs to the target audience of a buyer's data request, without revealing any additional information.

\end{abstract}

%---------------------------------------

%---------------------------------------
\section{Introduction}
%---------------------------------------

In this study, we are interested in the problem of trading real-world private information using only cryptographic protocols and a public blockchain to guarantee honest transactions.
Private information in this context refers to attributes or events associated with a single individual (person or organization). % , or its relationship with other individuals. 

In \cite{Travizano2018Wibson} the authors introduced the notion of a decentralized Privacy-Preserving Data Marketplace (\dPDM).
A decentralized Data Marketplace (\dDM) is a Data Marketplace (\DM) with no central authority, no central data repository and no central funds repository.
Additionally, a {\dPDM}  allows users to sell private information, while providing them privacy guarantees such as:
\begin{itemize}
\item Participant anonymity: the identities of the Sellers and Buyers are not revealed without their consent. In particular, the identity of the Data Seller is not revealed to the Data Buyer, without the consent of the Data Seller.
\item Transparency over Data usage: the Data Seller always has visibility on how his Data is used by the Buyer.
\end{itemize}

Here we consider three types of agents interacting in a {\dPDM} both privately (through end-to-end communications) and publicly through a permissionless blockchain:
\begin{description}
\item[Data Seller:]
The Seller $\fraS$ is the owner and subject of the private information that will be traded. He decides when and if his data is sold.

\item[Data Buyer:]
The Buyer $\fraB$ is interested in acquiring information from Sellers, provided that the information meets the Buyer's quality requirements.

\item[Notary:]
The Notary $\fraN$ has the means to validate information associated with Sellers.  He is trusted by Buyers to certify the precision and quality of the data traded.   
The Notary is the only public player with a formal track record and a public reputation. 

\end{description}

The information traded is collected mainly outside of the blockchain. The Notary will typically access the data as part of its business operations with the Sellers. The Buyer understands the value associated with the privileged position of the Notary and knows about the incentives aligned with its reputation.

The design and price of information in data markets is an active field of study~\cite{admati1990direct,bergemann2018markets,bergemann2018design,bergemann2015limits}. In the marketplace considered here, Data Sellers are able to participate in a decentralized marketplace that provides both financial incentives and control over their personal information~\cite{fernandez2020wibson}.
In addition, the Wibson platform provides a mechanism for the secure exchange of digital goods~\cite{futoransky2019secure} and a gas efficient protocol named BatPay for the recurrent micropayment of tokens~\cite{batpay2020}.

%---------------------------------------
\section{Problem Statement}
%---------------------------------------

%---------------------------------------
\subsection{Precise Information Trading}
%---------------------------------------

We would like to construct a platform that allows the trading of personal data with very precise granularity.
This means that the amount of information gained by a Data Buyer $\fraB$ in each transaction $\fraT$ can be properly constrained and narrowed.

In addition, the Data Seller $\fraS$ should not gain any information about other Sellers participating in other similar transactions. The Notary that certifies the Seller’s Data should gain minimal information as well.

Under some circumstances, it is also required that the Seller should not learn about the Buyer’s search criteria. 

For example:
\begin{itemize}
\item The Buyer should only learn if the Seller matches its search criteria, and nothing else. 
\item If a Seller participates in different transactions with the same Buyer, it should not be able to distinguish if those operations belong to the same agent.
\item The Notary which certifies a Seller’s personal information should not learn additional information about the Seller.
\end{itemize}

Zero Knowledge (ZK) proof systems have been proposed~\cite{bitansky2013succinct,campanelli2017zero} which provide cryptographic tools for these issues. In particular, there are now practical universal solutions such as zk-SNARK and zk-STARK protocols for ZK proofs~\cite{ben2013snarks,sasson2014zerocash,costello2015geppetto,ben2018scalable}.
However, these ZK proof protocols are still too expensive in terms of computational and storage cost required to generate the proof and, to a lesser extent, to verify it. 
They are particularly expensive when the algorithms are run in mobile devices with restricted computing and storage capacities.

On the other side, MPC-based solutions like ZKBoo~\cite{jawurek2013zero,giacomelli2016zkboo} are inexpensive in terms of computational cost but their communication requirements render them unacceptable for our application.

%---------------------------------------
\subsection{Problem Illustration}
%---------------------------------------

Whenever a Data Buyer $\fraB$ specifies a search criterion that will be used to identify potential Sellers, each participant will need to prove that he is a match for the Data Request. But special care has to be taken in order to prevent additional information from leaking. We illustrate this problem with sample requests:
\begin{itemize}
\item Buyer $\fraB$: Looking for people age 20-30 
\item Seller $\fraS_1$:  “I am 25” 
\item Seller $\fraS_2$: “I am in the 20-30 range”
\end{itemize}

\begin{itemize}
\item Buyer $\fraB$: Looking for people who visited a.com, b.com or c.com 
\item Seller $\fraS_1$: “I visited a.com” 
\item Seller $\fraS_2$: “I visited one of those sites”
\end{itemize}

\begin{itemize}
\item Buyer $\fraB$: Looking for people living inside this particular area (polygon) 
\item Seller $\fraS_1$: “I live at (latitude, longitude)” 
\item Seller $\fraS_2$: “I live inside the polygon”
\end{itemize}

\begin{itemize}
\item Buyer $\fraB$: Looking for people with at least \$10k in the bank 
\item Seller $\fraS_1$: “I have \$23k” 
\item Seller $\fraS_2$: “I have at least \$10k” 
\end{itemize}

In these examples, Seller $\fraS_1$ is revealing more information than needed, whereas Seller $\fraS_2$ is revealing just the right amount of information required in order to participate in the data transaction.

The  protocol that we describe in the next sections will allow a Data Seller to respond to a Data Request without revealing any additional information.

%---------------------------------------
\subsection{A Naive Solution}
%---------------------------------------
 
A very simple solution can be constructed, wherein Data Buyers request the Notary for signed copies of the information they are looking for. On the other hand, the Data Sellers are consulted just to authorize the transactions.
 
This scenario has several problems:
\begin{itemize}
\item Operations depend on the availability and scalability of the Notaries.
\item Notaries have excessive control over the marketplace, and can exercise arbitrary blocking on requests from selected Buyers.
\item Notaries would learn a lot about the Data Buyer’s search criteria.
\item Notaries would gain a lot of additional information from other Notaries.
\end{itemize}

A better solution should have Notaries issuing certificates prior to the Buyer’s requests. These certificates could be stored by Sellers and used whenever needed to complete orders generated by Buyers.
The following protocols implement these ideas in the context of a {\dPDM}.

%---------------------------------------
\section{Building Blocks}
%---------------------------------------

%---------------------------------------
\subsection{Match Criterion as Function}
%---------------------------------------

The general idea is the following:
\begin{enumerate}
\item The Notary $\fraN$ will issue a function $F_\fraS$ associated with each Seller $\fraS$. This function will take a description of a criterion as argument and return true or false depending on whether the Seller matches the criterion:
$$ 
F_\fraS(X) \in \{\mathsf{true}, \mathsf{false}\} . 
$$
 
\item The Notary will sign the function:
$$
			\mbox{Sign}_{Notary}(F_\fraS).
$$
\item The Buyer $\fraB$ specifies a criterion $X_0$. 

\item If the Seller matches the search criterion, he shows that $F_\fraS(X_0) = \mathsf{true}$, and that $F_\fraS(X_0)$ is a valuation of the same function $F_\fraS$ which was signed by the Notary.

\item For any other $X \neq X_0$, the Buyer does not learn whether $F_\fraS(X)$ is true or false.
\end{enumerate}

For example the following functions represent evaluations of different Buyer's criteria for a given Data Seller $\fraS$:
\begin{description}
\item[\normalfont -- $ \mathsf{ageInRange}_{\fraS}(\mbox{minAge}, \mbox{maxAge}) \rightarrow \{\mathsf{true}, \mathsf{false}\} $] ~\\
Returns true if the age of the seller $\fraS$ is within the specified minAge and maxAge. 
 
\item[\normalfont -- $ \mathsf{visitedAnySite}_\fraS(\mbox{siteList}) \rightarrow \{\mathsf{true}, \mathsf{false}\} $] ~\\
Returns true if the browsing history of the seller $\fraS$ contains any of the domains listed as a parameter.

\item[\normalfont -- $ \mathsf{houseInPoly}_\fraS(\mbox{polygon}) \rightarrow \{\mathsf{true}, \mathsf{false}\} $] ~\\
Returns true if the home address of the seller $\fraS$ is within the specified polygon. 

\item[\normalfont -- $ \mathsf{bankBalanceAtLeast}_\fraS(\mbox{minBalance}) \rightarrow \{\mathsf{true}, \mathsf{false}\} $] ~\\
Returns true if the bank balance of the seller $\fraS$ is at least minBalance.

\end{description}

%---------------------------------------
\subsection{Ordered Binary Decision Diagrams}
%---------------------------------------

We are going to represent the selected functions as binary decision diagrams (also known as branching programs).

An \emph{Ordered Binary Decision Diagram} (OBDD) is a binary tree that represents a function. Its value can be obtained by traversing a particular path from root to leaf.  The input argument is split into bits, where each bit in order is used node by node to decide whether to continue left or right. Finally, on the selected leaf, the output of the function is read.

In addition, the tree could be compressed by representing similar subtrees (different nodes on the same level that produce equivalent results for every input combination) as a single node. This transforms the tree into a \emph{Direct Acyclic Graph} (DAG). But the semantic evaluation of the graph as a function is retained.

\begin{figure}[ht]
\centering
\begin{subfigure}[b]{0.22\textwidth}
\centering
	{\includegraphics[width=.75\textwidth]{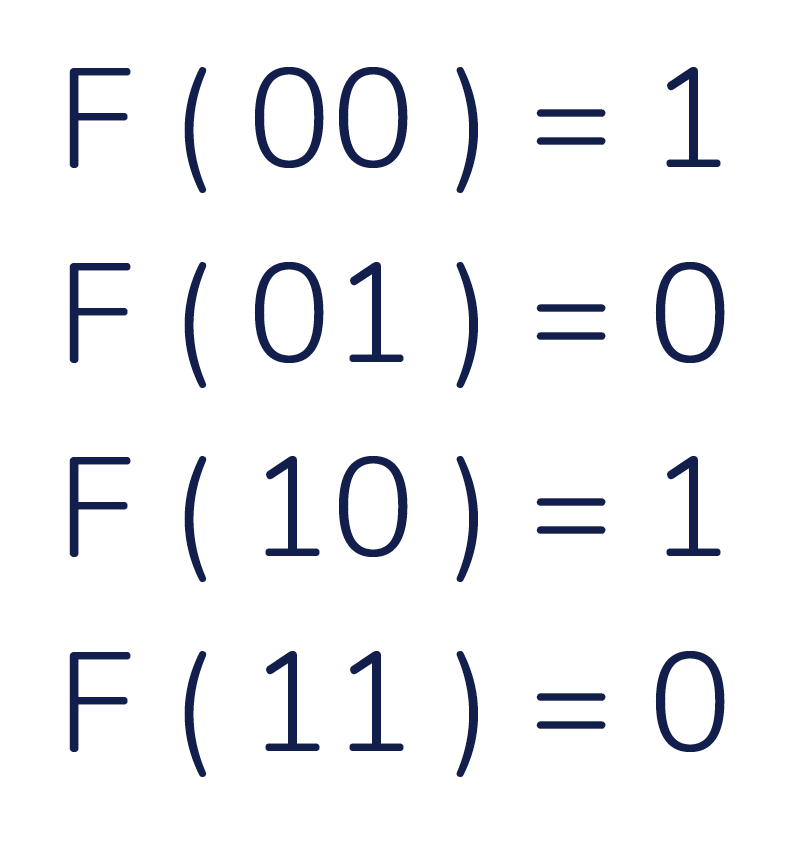}} 
	\caption{Function $F$}
\end{subfigure}
\begin{subfigure}[b]{0.44\textwidth}
\centering
	{\includegraphics[width=.85\textwidth]{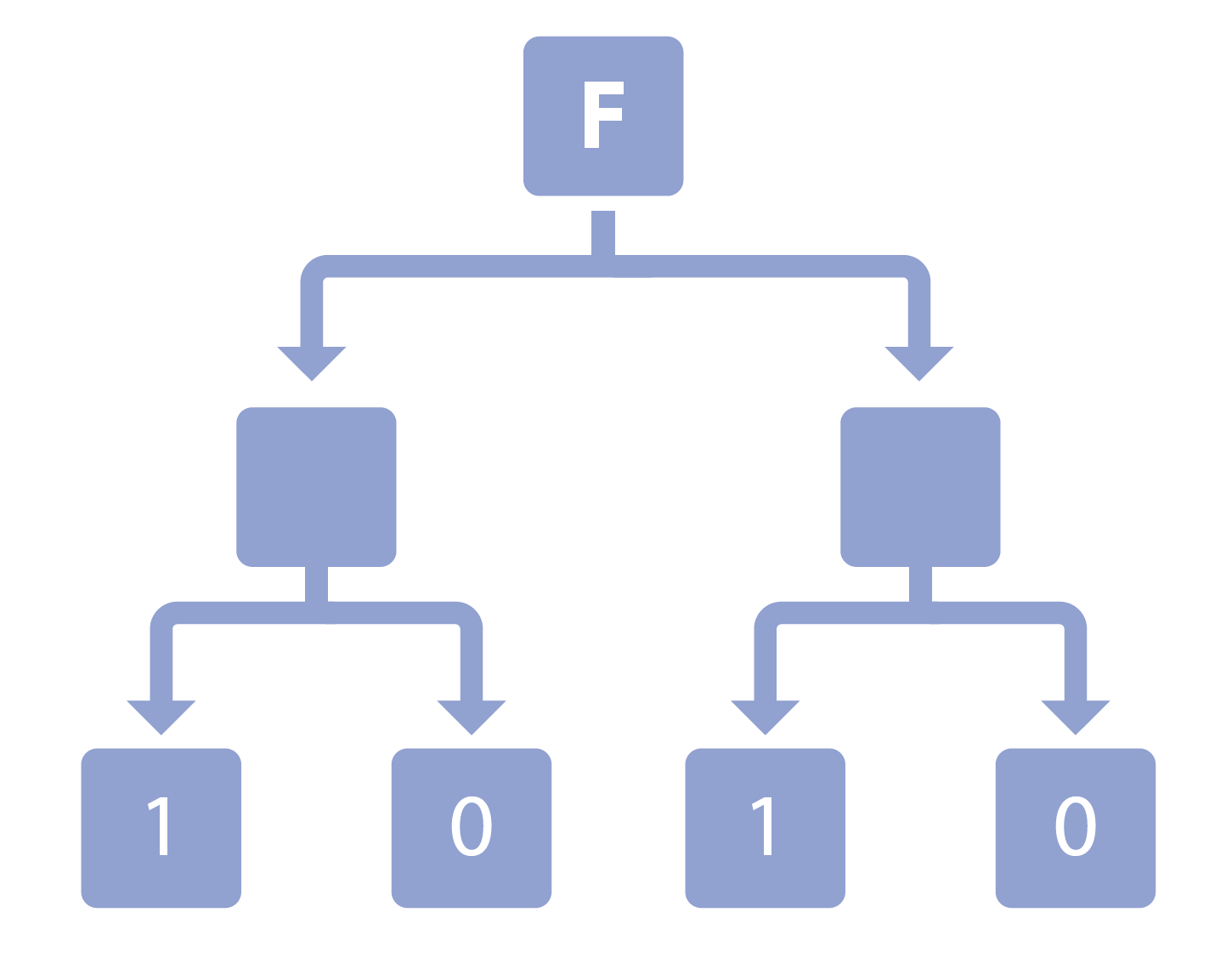}} 
	\caption{OBDD for $F$}
\end{subfigure}
\begin{subfigure}[b]{0.27\textwidth}
\centering
	{\includegraphics[width=.65\textwidth]{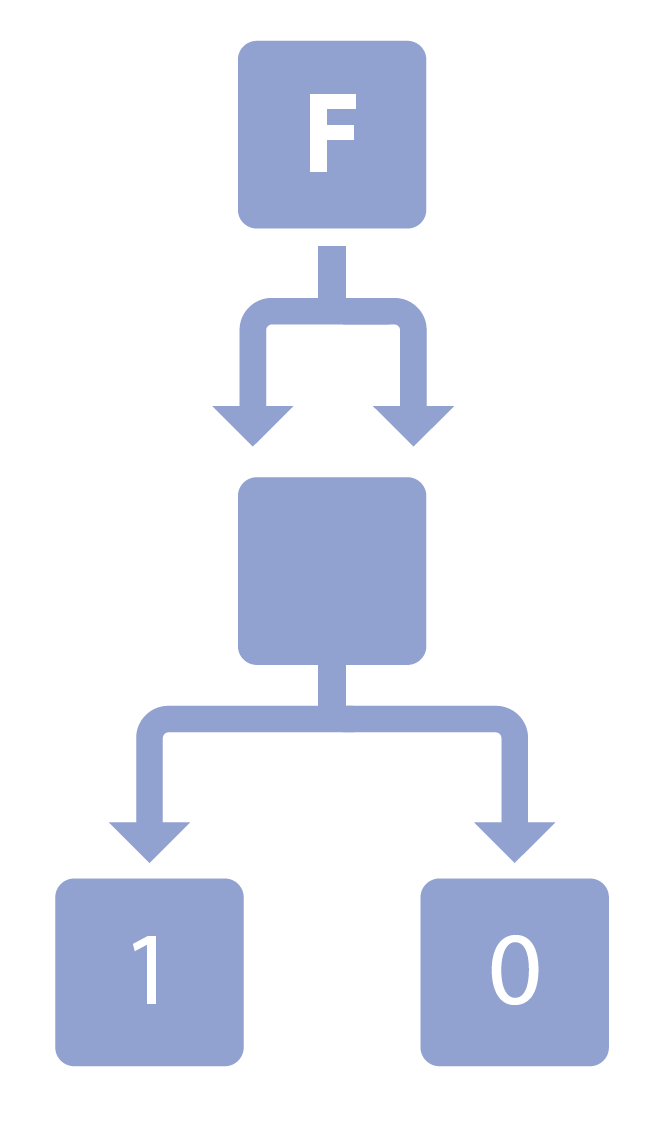}}
	\caption{Compressed OBDD}
\end{subfigure}
\caption{Example representation of (a) function $F$, (b) Ordered Binary Decision Diagram (OBDD) for $F$ and (c) compressed OBDD for $F$.}
\label{fig:obdd}
\end{figure}

This compression enables the representation of functions with larger input sizes. 
Not all functions can be efficiently expressed in this way, but we found a nice subset of interesting criterion predicates that work well, including:
\begin{itemize}
\item Integer and fixed-point ranges.
\item Strings and Finite state automaton.
\item Matching with string-sets with wildcards.
\item Spatial coordinates, including approximation of convex polygons, and union of convex polygons.
\item Simple logical expressions.

\end{itemize}

%---------------------------------------
\section{Creating a WibsonTree}
%---------------------------------------

In this section we introduce  the \emph{WibsonTree} primitive, which enables the creation of cryptographic commitments for a function $F$, represented as an OBDD. Later, any single path can be opened and verified against the commitment without exposing additional information.

The data structure resembles a sparse Merkle-tree, which includes per-level blinding keys. We use $\mathsf{sha256}$ as the hash function in our implementations, but any other hash primitive can be used.

Given a function $F$ with inputs in ${\Z_2}^n$, we create $n$ pairs of random keys: 
$$
\{ L_i,  R_i \} \mbox{ for } 0 \leq i < n. 
$$

A keyed crypto-secure Pseudo-Random Number Generator (PRNG) is used to derive these keys based on a single random seed. 

We will then assign a hash $H_{node}$ to every node in the tree.

For every \textbf{leaf node} (output node), its hash is:
$$
		H_{node} = \Hash ( \mbox{Result}_{node}).
$$

For any \textbf{non-leaf node} at level $i$, whose children hashes are $Child_L$ and $Child_R$, its hash can be calculated as:
$$
H_{node} = \Hash \left( \Hash (Child_L \, || \, L_i) \; || \; \Hash(Child_R \, || \, R_i) \right)
$$

\begin{figure}[htbp]
\begin{center}
{\includegraphics[width=0.65\textwidth]{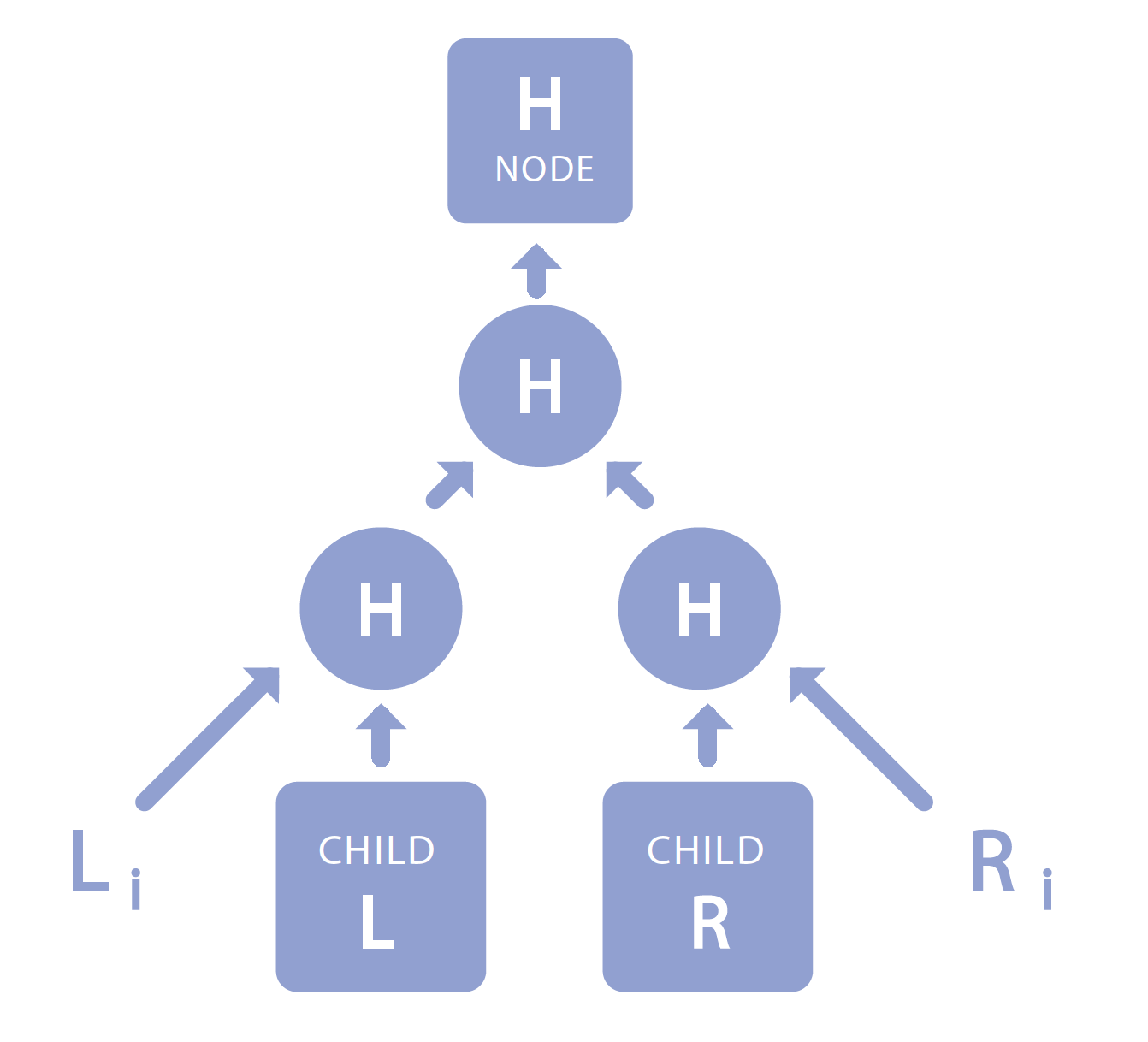} 
}
\caption{Assigning a hash to a non-leaf node while creating a WibsonTree.}
\label{fig:wibsontree2}
\end{center}
\end{figure}

In addition, an auxiliary value is calculated for each node:
$$
\AUX_{node} = \Hash(Child_L \, || \, L_i) \oplus \Hash(Child_R \, || \, R_i)
$$
The hash for the root node ($H_{root}$) represents the binding commitment for the tree and the associated function.

%---------------------------------------
\subsection{Opening a Commitment}
%---------------------------------------

In general, we would like to open a commitment for a single point of the function, while hiding the rest. In the unusual case that the whole function needs to be revealed, the complete set of keys $ \{ L_i, R_i \}_{0 \leq i < n} $ can be shared.

The more usual and interesting case is when a single input $X$ has been specified. In this case, a witness can be created as follows:

	First, for each bit $i$ in $X$, depending on the bit value $X_i$, $L_i$ or $R_i$ are revealed. That is, the binary representation of the input is used to alternatively select one of the two keys available for each level:
	\begin{align*}
	\mbox{If } X_i &= 0  \mbox{:  reveal  } L_i \\
	\mbox{If } X_i &=1 \mbox{:  reveal  } R_i 
	\end{align*}

	In addition, for every node $j$ in the path implied by $X$, from root to every intermediate node except the result leaf node, $\AUX_j$ is revealed. 

For the leaf node, the result is appended to the witness.

%---------------------------------------
\subsection{Verifying a Commitment}
%---------------------------------------

To verify a commitment opening, we traverse the path backwards, from leaf to root, using keys and $\AUX$ information to reconstruct node hashes, until the root hash has been calculated. If the opening is correct, the root hash should match the original commitment’s hash.

For a \textbf{leaf node}: 
$$
	H_{node} = \Hash(\mbox{Result}_{node})
$$
For a \textbf{non-leaf node} at level $i$ with known child’s hash $c$:
$$
	\mbox{If } X_i = 0: H_{node} = \Hash(  \Hash( c \, || \, L_i ) \; || \; [ \Hash( c \, || \, L_i )  \oplus \AUX_{node} ] ) 
$$
$$
	\mbox{If } X_i = 1: H_{node} = \Hash(  [ \Hash( c \, || \, R_i )  \oplus \AUX_{node} ] \; || \; \Hash( c \, || \, R_i ) )
$$

%---------------------------------------
\subsection{WibsonTree Efficiency}
%---------------------------------------

WibsonTrees are created out of Ordered Binary Decision Diagrams. With fixed L/R keys, the calculation of the root hash is not dependent on whether or how the tree has been compressed. An efficient representation of the tree as a DAG will result in an efficient calculation of the root hash.
There are many powerful algorithms to produce optimized representations of a branching program.    

In addition, the witness generated while opening a commitment is quite compact: 1 hash and 1 key per tree level minus 1. Note that the witness is not dependent on compression, so information about the redundancy of the function is not revealed.

Moreover, the only primitive required to create a Wibson tree is the hash function, which has much less computational cost and complexity than zk-SNARK~\cite{ben2013snarks}.

%---------------------------------------
\section{Conclusion}
%---------------------------------------

In this paper, we presented a cryptographic primitive called \emph{WibsonTree}
designed to preserve users' privacy
by allowing them to demonstrate predicates on their personal attributes, 
without revealing the values of those attributes.

This primitive is presented in the context of a decentralized Privacy-preserving Data Marketplace (\dPDM)
such as the Wibson marketplace~\cite{fernandez2020wibson}.
In a {\dPDM} there are three types of agents: Data Buyers, Data Sellers and Notaries.
 
The \emph{WibsonTree} protocol that we presented is an efficient cryptographic primitive that enables the exchange of private information while preserving the Data Seller's privacy by using proof-kits previously issued by the Notary. 
By using our primitive, a Data Seller can efficiently prove (with small computing requirements) that he/she belongs to the target audience of a Buyer's data request, without revealing any additional information.

%----------------------------------------------------
% \renewcommand{\arraystretch}{1.3}

\setlength{\parskip}{1pt}
\renewcommand{\arraystretch}{1.0}

\bibliographystyle{unsrt}
\bibliography{../wibson}

\begin{thebibliography}{10}

\bibitem{Travizano2018Wibson}
Matias Travizano, Carlos Sarraute, Gustavo Ajzenman, and Martin Minnoni.
\newblock Wibson: A decentralized data marketplace.
\newblock In {\em Proceedings of SIGBPS 2018 Workshop on Blockchain and Smart
  Contract}, 2018.

\bibitem{admati1990direct}
Anat~R Admati and Paul Pfleiderer.
\newblock Direct and indirect sale of information.
\newblock {\em Econometrica: Journal of the Econometric Society}, pages
  901--928, 1990.

\bibitem{bergemann2018markets}
Dirk Bergemann, Alessandro Bonatti, et~al.
\newblock Markets for information: An introduction.
\newblock Technical report, Cowles Foundation for Research in Economics, Yale
  University, 2018.

\bibitem{bergemann2018design}
Dirk Bergemann, Alessandro Bonatti, and Alex Smolin.
\newblock The design and price of information.
\newblock {\em American Economic Review}, 108(1):1--48, 2018.

\bibitem{bergemann2015limits}
Dirk Bergemann, Benjamin Brooks, and Stephen Morris.
\newblock The limits of price discrimination.
\newblock {\em American Economic Review}, 105(3):921--57, 2015.

\bibitem{fernandez2020wibson}
Daniel Fernandez, Ariel Futoransky, Gustavo Ajzenman, Matias Travizano, and
  Carlos Sarraute.
\newblock Wibson protocol for secure data exchange and batch payments.
\newblock {\em arXiv:2001.08832}, 2020.

\bibitem{futoransky2019secure}
Ariel Futoransky, Carlos Sarraute, Ariel Waissbein, Daniel Fernandez, Matias
  Travizano, and Martin Minnoni.
\newblock Secure exchange of digital goods in a decentralized data marketplace.
\newblock In {\em Proceedings of the 2019 Argentine Symposium on Big Data
  (AGRANDA)}, pages 38--44, 2019.

\bibitem{batpay2020}
Hartwig Mayer, Ismael Bejarano, Daniel Fernandez, Gustavo Ajzenman, Nicolas
  Ayala, Nahuel Santoalla, Carlos Sarraute, and Ariel Futoransky.
\newblock {B}at{P}ay: a gas efficient protocol for the recurrent micropayment
  of {ERC20} tokens.
\newblock {\em arXiv:2002.02316}, 2020.

\bibitem{bitansky2013succinct}
Nir Bitansky, Alessandro Chiesa, Yuval Ishai, Omer Paneth, and Rafail
  Ostrovsky.
\newblock Succinct non-interactive arguments via linear interactive proofs.
\newblock In {\em Theory of Cryptography}, pages 315--333. Springer, Berlin,
  Heidelberg, 2013.

\bibitem{campanelli2017zero}
Matteo Campanelli, Rosario Gennaro, Steven Goldfeder, and Luca Nizzardo.
\newblock Zero-knowledge contingent payments revisited: Attacks and payments
  for services.
\newblock In {\em Proceedings of the 2017 ACM SIGSAC Conference on Computer and
  Communications Security}, pages 229--243. ACM, 2017.

\bibitem{ben2013snarks}
Eli Ben-Sasson, Alessandro Chiesa, Daniel Genkin, Eran Tromer, and Madars
  Virza.
\newblock Snarks for c: Verifying program executions succinctly and in zero
  knowledge.
\newblock In {\em Advances in Cryptology--CRYPTO 2013}, pages 90--108.
  Springer, 2013.

\bibitem{sasson2014zerocash}
Eli Ben-Sasson, Alessandro Chiesa, Christina Garman, Matthew Green, Ian Miers,
  Eran Tromer, and Madars Virza.
\newblock Zerocash: Decentralized anonymous payments from bitcoin.
\newblock In {\em 2014 IEEE Symposium on Security and Privacy (SP)}, pages
  459--474. IEEE, 2014.

\bibitem{costello2015geppetto}
Craig Costello, C{\'e}dric Fournet, Jon Howell, Markulf Kohlweiss, Benjamin
  Kreuter, Michael Naehrig, Bryan Parno, and Samee Zahur.
\newblock Geppetto: Versatile verifiable computation.
\newblock In {\em 2015 IEEE Symposium on Security and Privacy (SP)}, pages
  253--270. IEEE, 2015.

\bibitem{ben2018scalable}
Eli Ben-Sasson, Iddo Bentov, Yinon Horesh, and Michael Riabzev.
\newblock Scalable, transparent, and post-quantum secure computational
  integrity.
\newblock {\em Cryptol. ePrint Arch., Tech. Rep}, 46:2018, 2018.

\bibitem{jawurek2013zero}
Marek Jawurek, Florian Kerschbaum, and Claudio Orlandi.
\newblock Zero-knowledge using garbled circuits: how to prove non-algebraic
  statements efficiently.
\newblock In {\em Proceedings of the 2013 ACM SIGSAC conference on Computer \&
  communications security}, pages 955--966. ACM, 2013.

\bibitem{giacomelli2016zkboo}
Irene Giacomelli, Jesper Madsen, and Claudio Orlandi.
\newblock Zkboo: Faster zero-knowledge for boolean circuits.
\newblock In {\em USENIX Security Symposium}, pages 1069--1083, 2016.

\end{thebibliography}

%----------------------------------------------------

\end{document}